\documentclass[prd,twocolumn,showpacs,amsmath,amssymb]{revtex4-1}
\usepackage{amssymb}
\usepackage{amsfonts}
\usepackage{overpic,graphicx}
\usepackage{keyval,graphicx}
\usepackage{textcomp,wasysym}
\usepackage[perpage,symbol]{footmisc}%
\usepackage{lineno}
\usepackage[dvipdfm,CJKbookmarks=true, colorlinks=true,
linkcolor=blue, urlcolor=blue,citecolor=blue]{hyperref}
\usepackage[normalem]{ulem}
\usepackage{lineno}

\begin{document}

\title{\boldmath  Observation of $J/\psi\to \gamma\eta\pi^{0}$ }


\author{
\begin{small}
\begin{center}
M.~Ablikim$^{1}$, M.~N.~Achasov$^{9,e}$, S. ~Ahmed$^{14}$, X.~C.~Ai$^{1}$, O.~Albayrak$^{5}$, M.~Albrecht$^{4}$, D.~J.~Ambrose$^{45}$, A.~Amoroso$^{50A,50C}$, F.~F.~An$^{1}$, Q.~An$^{47,a}$
% ...
\\
\vspace{0.2cm}
(BESIII Collaboration)\\
\vspace{0.2cm} {\it
$^{1}$ Institute of High Energy Physics, Beijing 100049, People's Republic of China\\
% ...
$^{4}$ Bochum Ruhr-University, D-44780 Bochum, Germany\\
$^{5}$ Carnegie Mellon University, Pittsburgh, Pennsylvania 15213, USA\\
% ...
$^{9}$ G.I. Budker Institute of Nuclear Physics SB RAS (BINP), Novosibirsk 630090, Russia\\
% ...
$^{14}$ Helmholtz Institute Mainz, Johann-Joachim-Becher-Weg 45, D-55099 Mainz, Germany\\
% ...
$^{45}$ University of Rochester, Rochester, New York 14627, USA\\
% ...
$^{47}$ University of Science and Technology of China, Hefei 230026, People's Republic of China\\
% ...
$^{50}$ (A)University of Turin, I-10125, Turin, Italy; (B)University of Eastern Piedmont, I-15121, Alessandria, Italy; (C)INFN, I-10125, Turin, Italy\\
% ...
$^{a}$ Also at State Key Laboratory of Particle Detection and Electronics, Beijing 100049, Hefei 230026, People's Republic of China\\
% ...
$^{e}$ Also at the Novosibirsk State University, Novosibirsk, 630090, Russia\\
% ...
}\end{center}
\end{small}
}

\begin{abstract}
We present the first study of the process $J/\psi \rightarrow \gamma\eta\pi^{0}$ using
$(223.7\pm1.4)\times10^{6}$ $J/\psi$ events accumulated with the BESIII detector at the BEPCII facility.
The branching fraction for $J/\psi \rightarrow \gamma\eta\pi^{0}$ is measured to be
$\mathcal{B}(J/\psi \rightarrow \gamma\eta\pi^{0}) =(2.14\pm0.18(stat)\pm0.25(syst))\times10^{-5}$.
With a Bayesian approach, the upper limits of the branching fractions
$\mathcal{B}(J/\psi \rightarrow \gamma a_{0}(980),a_0(980)\rightarrow\eta\pi^0)$ and
$\mathcal{B}(J/\psi \rightarrow \gamma a_{2}(1320),a_2(1320)\rightarrow\eta\pi^0)$ are determined to be
$2.5\times10^{-6}$ and $6.6\times10^{-6}$ at the 95\% confidence
level, respectively.  All of these measurements are given for the first time.
\end{abstract}

\maketitle
\newpage

\section{Introduction}\label{intro}

The nature of the lightest scalar meson nonet has been a hot topic in hadron physics for many years~\cite{ref:pdg}.
In particular, the nature of the isovector $a_0(980)$ is still not understood.
It is interpreted by theorists to be a $q\bar{q}$ state with a possible admixture of a $K\bar{K}$ bound-state component due to the proximity to the $K\bar{K}$ threshold~\cite{ref:pdg, ref:baru, ref:a0f0}.
The $a_0(980)$ mass is known to be about 980 MeV and the dominant decay mode is $a_0(980)\to \eta \pi$.
The radiative decay of the $J/\psi$ to the enigmatic scalar meson $a_0(980)$  will provide useful information on the nature of $a_0(980)$ state~\cite{jpsidecay:1989, ref:theory}.
Especially, in Ref.~\cite{ref:theory},  the predicted branching fraction is ${\cal B}(J/\psi \to \gamma a_0(980))= (3.1 \pm 1.5)\times 10^{-3}$ based on the factorization of mixing and effective coupling constants. Therefore,  search for production of the neutral
$a_0(980)$ in the isospin-violating decay $J/\psi \to \gamma \eta \pi^0$ will discriminate between different models~\cite{jpsidecay:1989, ref:theory}.

The radiative $J/\psi$ decays with the total isospin of the hadronic final state $I=0$, such as
$J/\psi\rightarrow\gamma \pi\pi$ or $J/\psi\rightarrow\gamma \eta \eta$, have been studied by
previous experiments~\cite{ref:liucy,ref:previous,ref:zhuyc,ref:bes1,ref:gpi0pi0},
while only a few processes with isotriplet hadronic final states, such as $J/\psi\rightarrow\gamma \pi^{0}$ and $J/\psi\rightarrow\gamma \pi^{0}\pi^{0}\pi^{0}$,
have been measured~\cite{ref:isospin1,ref:isospin11}.
It is therefore of interest to study the isospin violating decay $J/\psi\rightarrow\gamma \eta \pi^0$, which can be used to test charmonium  decay dynamics~\cite{ref:theory}.

In this paper, we present the first study of the decay $J/\psi\to\gamma\eta\pi^{0}$
based on a sample of $(223.7\pm1.4)\times 10^6$ $J/\psi$
events~\cite{ref:jpsitotnumber}, collected by the Beijing
Spectrometer (BESIII) located at the Beijing Electron Positron Collider
(BEPCII).

\section{BESIII DETECTOR AND DATA SAMPLES}\label{BESIIIBEPC}

The accelerator BEPCII and the BESIII detector~\cite{bes3} are major upgrades of the BESII experiment
at the BEPC accelerator~\cite{bes2,bes2-p2} for studies of hadron
spectroscopy, charmonium physics, and $\tau$-charm physics~\cite{bes3phys}. The BESIII detector
with a geometrical acceptance of 93\% of 4$\pi$ consists of the
following main components: (1)~a small-cell
main drift chamber~(MDC) with 43 layers used to track charged particles. The average
single-wire resolution is 135~$\mu$m, and the momentum
resolution for 1~GeV/$c$ charged particles in a 1~T magnetic
field is 0.5\%. (2)~a time-of-flight system~(TOF) used for
particle identification.  It is composed of a barrel made of
two layers, each consisting of 88 pieces of 5~cm thick and 2.4~m long plastic
scintillators, as well as two end caps with 96 fan-shaped,
5~cm thick, plastic scintillators in each end cap.
The time resolution is 80~ps in the barrel and 110~ps in the
end caps, providing a $K/\pi$ separation of more than
2$\sigma$ for momenta up to about 1.0~GeV/$c$.
(3)~an electro-magnetic calorimeter~(EMC) used to measure photon energies.
The EMC is made of 6240 CsI~(Tl) crystals arranged in a cylindrical
shape (barrel) plus two end caps. For 1.0~GeV photons, the
energy resolution is 2.5\% in the barrel and 5\% in the end caps,
and the position resolution is 6~mm in the barrel and
9~mm in the end caps.  (4)~a muon counter made of resistive plate chambers
arranged in 9 layers in the barrel and 8 layers in the end caps, which
is incorporated into the iron flux return yoke of the superconducting
magnet. The position resolution is about 2~cm.

The event selection optimization, efficiency estimation, and background evaluation are performed
are performed through Monte Carlo (MC) simulations, for which the
{\sc GEANT}4-based~\cite{geant4} MC simulation package {\sc BOOST}~\cite{sim-boost} is used.
The {\sc BOOST} software incorporates the geometric and material description of the BESIII detector components,
the detector response and digitization models, and detector running conditions and performance.
The production of the $J/\psi$ resonance is simulated with the MC event generator {\sc KKMC}~\cite{sim-kkmc,sim-kkmc2},
while known decay modes are generated with {\sc EVTGEN}~\cite{sim-evtgen,ref:sim}, with branching
fractions set to world average values from the Particle Data Group (PDG)~\cite{ref:pdg}.
The {\sc LUNDCHARM}~\cite{sim-lundcharm} model is used for the remaining, unknown decays.
A sample of 200$\times 10^6$ generic  $J/\psi$ decay events (named inclusive
MC sample thereafter) is used to study potential backgrounds.
A sample of $10^5$ exclusive MC signal events $J/\psi\to\gamma\eta\pi^0\to 5\gamma$ is generated uniformly in phase space.
For additional signal studies, samples of $10^5$ exclusive $J/\psi\to \gamma a_0(980), a_0(980)\to \eta\pi^{0}$ and $J/\psi\to \gamma a_2(1320), a_2(1320)\to \eta\pi^{0}$
MC events are generated with angular dependence in the $\eta$ and $\pi^0$ distributions based on experimental information~\cite{sim-evtgen,ref:sim}.
For further background studies, we use $10^5$ exclusive MC events for each of the following processes:
$J/\psi \to \eta\omega (\eta \to \gamma\gamma, \omega \to \gamma\pi^{0})$, $J/\psi \to \eta\phi (\eta \to \gamma\gamma, \phi \to \gamma\pi^{0})$, $J/\psi \to \gamma\eta' (\eta'\to 2\pi^{0}\eta$ or $\eta' \to \gamma\omega)$.
All exclusive samples listed previously are generated without consideration of angular dependence in phase space.

\section{Event selection}\label{Event_select}
The $J/\psi \to \gamma\eta\pi^{0}$ decays, with subsequent decays $\eta \to \gamma\gamma$
and $\pi^{0} \to \gamma\gamma$, have a topology of five photons in the final state.
To select signal candidates, we require at least five photons and no reconstructed charged particles in an event.
The photon candidates are required to have at least 25 MeV deposited energy
in barrel region ($|\cos\theta| < 0.8$) of the EMC, while 50 MeV are required in the end cap regions
($ 0.86 < |\cos\theta| < 0.92$), where $\theta$ is the polar angle of the electromagnetic shower.
Timing information of the EMC is used to suppress electronic noise and energy depositions that are unrelated to the event.
Photon candidates within 50~ns relative to the most energetic shower are selected.

\begin{figure}[htbp]
\centering
\includegraphics[width=0.45\textwidth]{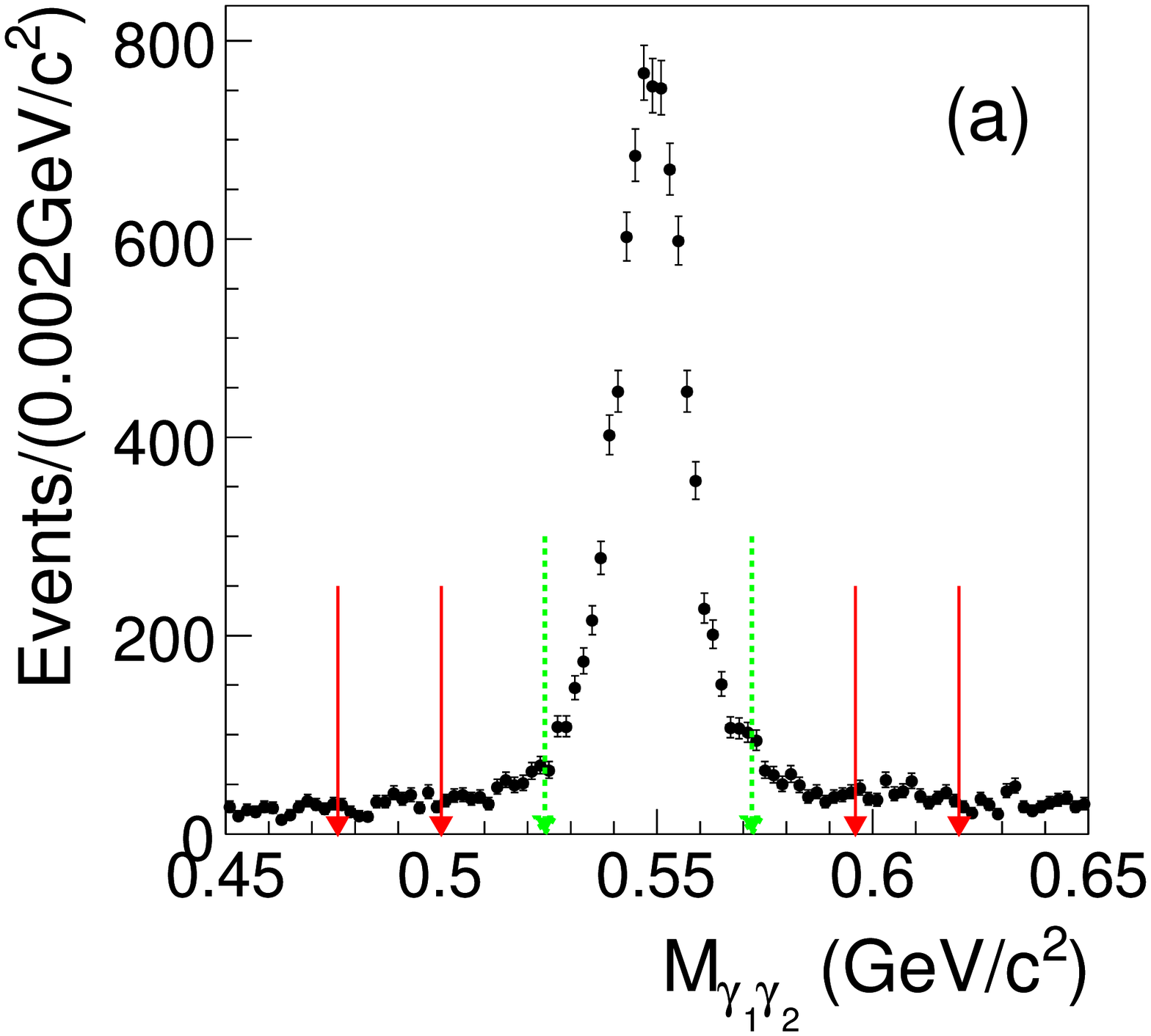}
\includegraphics[width=0.45\textwidth]{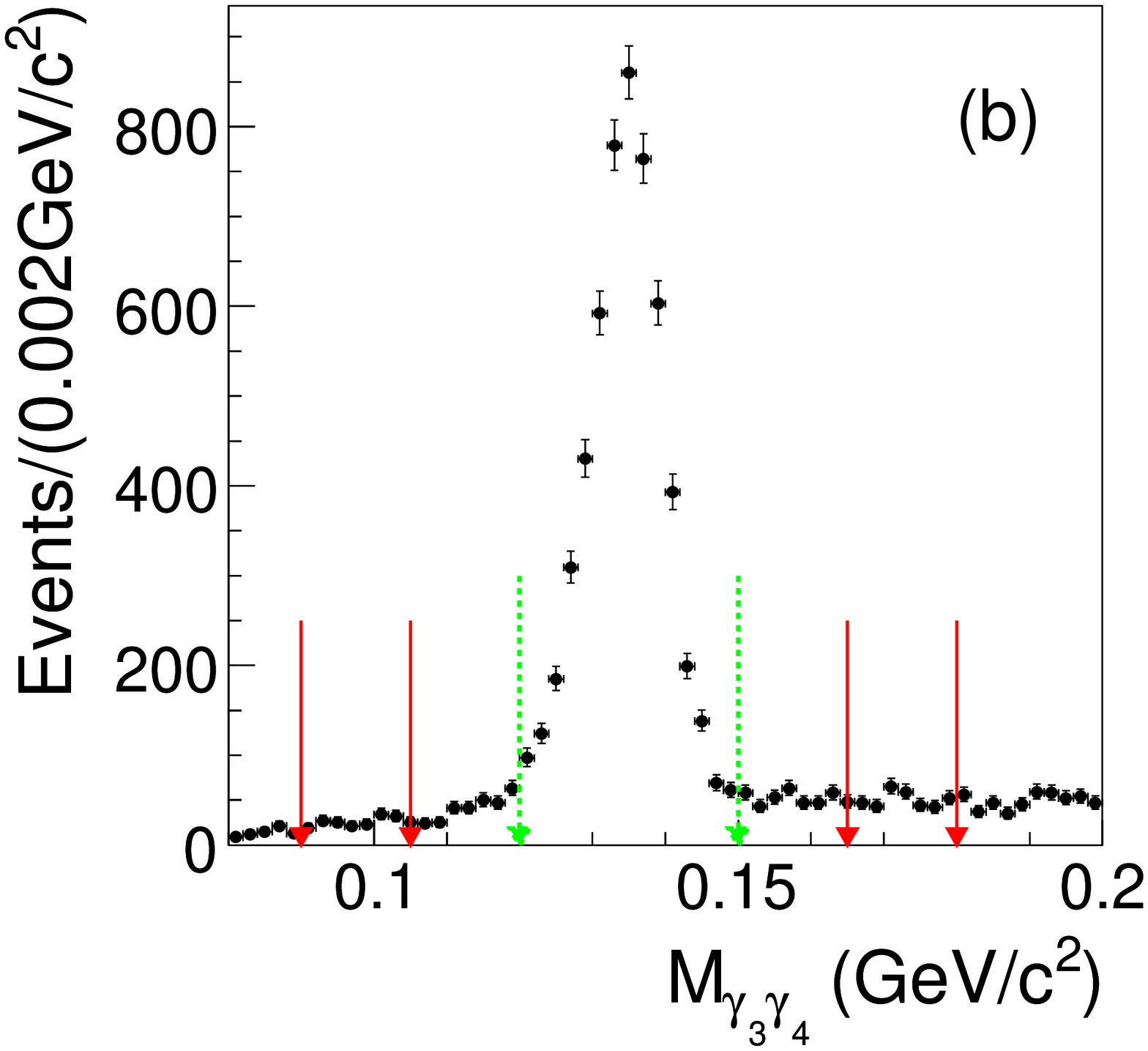}
\caption{Distributions of the $\gamma\gamma$ invariant mass from the $\eta$ (a) and $\pi^{0}$ (b) candidate decays. The arrows with dotted lines indicate the signal region, and the solid arrows indicate the sidebands.
\label{fig:mass_fit_data}}
\end{figure}

A four-constraint (4C) kinematic fit imposing energy-momentum
   conservation under the hypothesis $e^{+}e^{-} \to 5\gamma$ is performed, and
   $\chi^{2}_{4C} < 30$ is required.
All further selections are based on the four-momenta updated by the 4C fit.
The variable
   $\Delta = \sqrt{(M_{\gamma_{1}\gamma_{2}}-m_{\eta})^{2} + (M_{\gamma_{3}\gamma_{4}}-m_{\pi^{0}})^{2}}$ is used to identify which photons originate from the
   decays of $\eta$ and $\pi^0$, respectively;
 here, $M_{\gamma_i\gamma_j}$ is the invariant mass of two photons and $m_{\eta}$ ($m_{\pi^{0}}$) is the mass of $\eta$ ($\pi^{0}$) listed in PDG~\cite{ref:pdg}. We try all possible combinations of the five selected photons, and the one with the minimum  $\Delta$ is selected.
To suppress backgrounds
with two $\pi^{0}$ in the final state (\emph{e.g.}, $J/\psi \to \gamma\pi^{0}\pi^{0}$),
we define the variable $\Delta_{\pi^{0}}= \sqrt{(M_{\gamma_{1}\gamma_{2}}-m_{\pi^{0}})^{2} + (M_{\gamma_{3}\gamma_{4}}-m_{\pi^{0}})^{2}}$.
 An event is rejected if any combination of photons satisfies $\Delta_{\pi^{0}} < 0.05$ GeV$/c^{2}$.
The invariant mass spectra of the photon pairs from the $\eta$ and $\pi^{0}$ decays are shown in Fig.~\ref{fig:mass_fit_data}.
We fit a Gaussian function plus a
third order polynomial background to the mass spectra to obtain the mass resolution, which is determined to be 8 MeV$/c^{2}$ for the $\eta$ meson and 5 MeV$/c^{2}$ for the $\pi^{0}$.
The $\eta$ signal region is defined as $|M_{\gamma_{1}\gamma_{2}}- m_{\eta}| <0.024$~GeV/$c^{2}$.
The $\pi^{0}$ signal region is defined as $|M_{\gamma_{3}\gamma_{4}}- m_{\pi^{0}}| < 0.015$~GeV/$c^{2}$, and the $\pi^{0}$ sidebands are
defined as $0.030$~GeV/$c^{2} <|M_{\gamma_{3}\gamma_{4}}- m_{\pi^{0}}| < 0.045$~GeV/$c^{2}$.

\begin{figure}[htbp]
\centering
\includegraphics[width=0.45\textwidth]{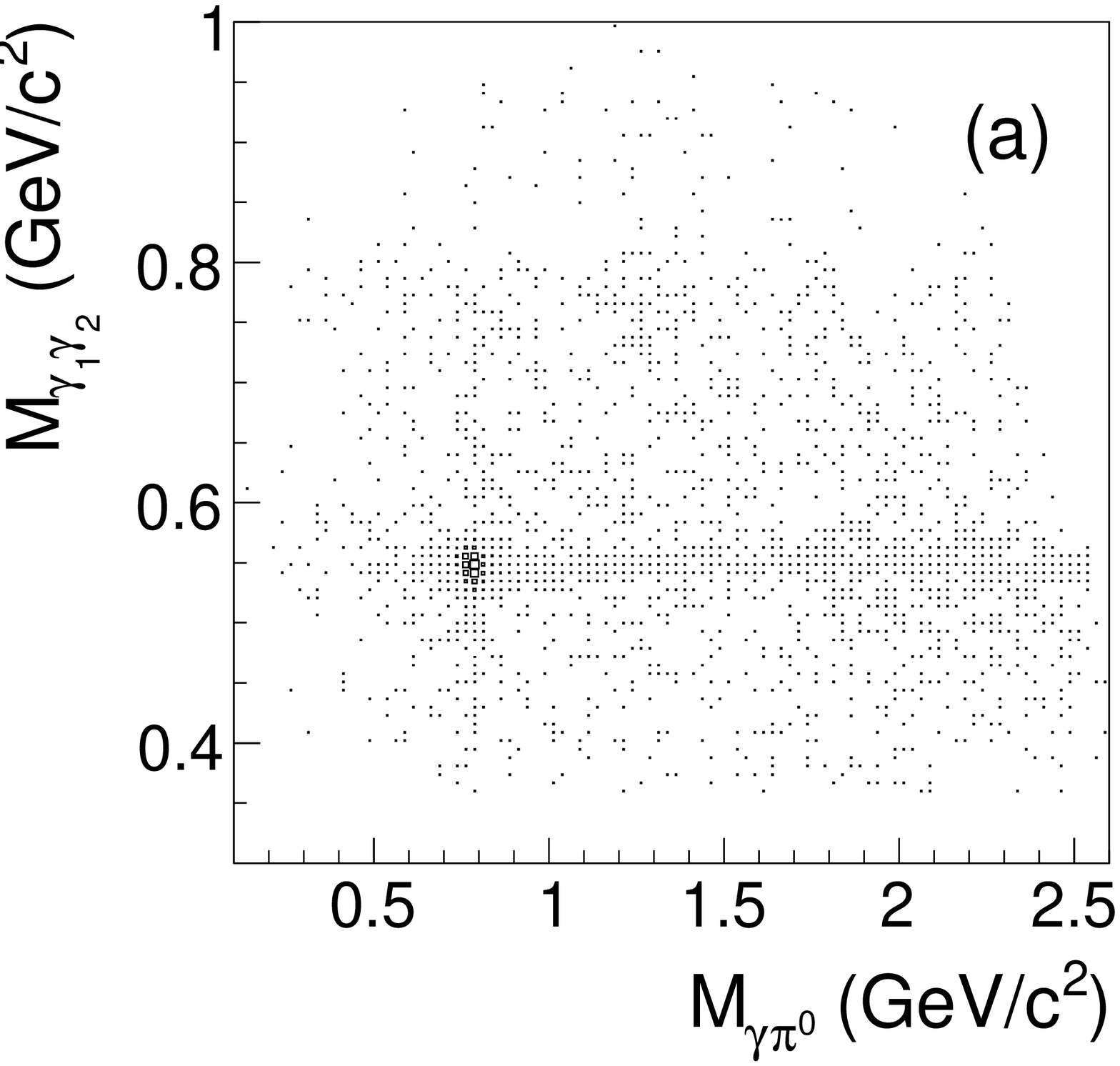}
\includegraphics[width=0.45\textwidth]{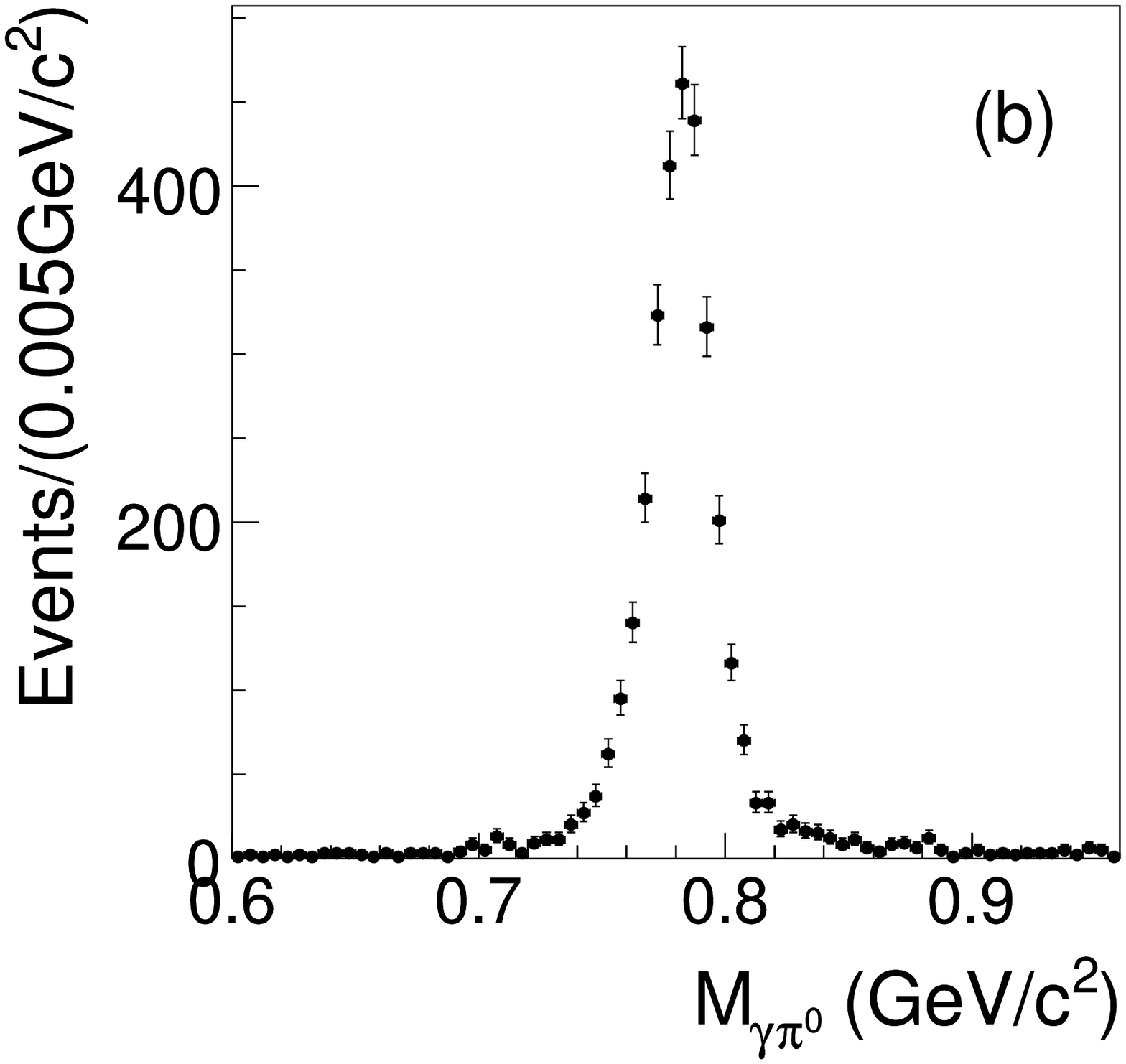}
\caption{(a) Scatter plot of $\gamma_1\gamma_2$ versus $\gamma\pi^{0}$ masses after selecting event candidates
with $\chi^{2}_{4C} < 30$ and $\Delta_{\pi^{0}} > 0.05$ GeV/$c^{2}$.
(b) The $\gamma\pi^{0}$ invariant mass spectrum after additional selection criteria are applied for photon-pair
candidates in the $\eta$ and $\pi^{0}$ signal regions.
\label{fig:scatter_omega}}
\end{figure}

The scatter plot of the invariant mass of the $\eta$ candidate versus that of $\gamma\pi^{0}$, obtained after applying above selection criteria,
is shown in Fig.~\ref{fig:scatter_omega}(a). A strong peak, which is associated with the background process from the production of $\omega$ mesons with the $\omega\to\gamma\pi^{0}$ final state, is visible in Fig.~\ref{fig:scatter_omega}(b).
The signature of the $\omega\to\gamma\pi^{0}$ decay is more evident from the invariant mass
spectrum shown in Fig.~\ref {fig:scatter_omega}(b), obtained after additionally selecting the $\eta$ and $\pi^{0}$ candidates.
To reject $\omega$ backgrounds, we require $|M_{\gamma\pi^{0}}-m_{\omega}| > 0.07$~GeV/$c^{2}$,
where $m_{\omega}$ is the nominal $\omega$ mass~\cite{ref:pdg}.

\section{Branching fraction and yield measurements}\label{br_getapi0}

After all selection criteria discussed in the previous section are applied,
we obtain event candidates for the decay $J/\psi \rightarrow \gamma\eta\pi^{0}$.
The potential background contribution is studied using both data and MC samples.
The background events from the data are selected using the $\pi^{0}$ sidebands, defined in Sec.~\ref{Event_select}.
In addition, the background events are studied with the inclusive $J/\psi$ MC sample;
the background events with the same final state are found to be from the $J/\psi\rightarrow \omega\eta ( \omega\rightarrow\gamma\pi^{0} )$ and
$J/\psi\rightarrow \phi\eta (\phi\rightarrow\gamma\pi^{0})$ decays. Apart from these two background channels,
other background contributions are found to be represented by the $\pi^{0}$ sidebands.

To scale the background events from the $\pi^0$ sideband regions to the signal region,
a normalization factor $f$ is defined as the ratio of the number of background events in
the $\pi^{0}$ signal region and in the $\pi^{0}$ sideband regions.
To obtain $f$, we fit to the $\pi^{0}$ mass spectrum
a combination of the $\pi^{0}$ signal shape, obtained from the exclusive signal MC, combined with a third order Chebychev polynomial to represent the background distribution.
The polynomial background is integrated in the signal region ($s_1$) and in the sideband regions ($s_2$) and
the normalization factor is found to be $f = \frac{s_1}{s_2} = 1.09$.

\begin{figure}[htbp]
\centering
\includegraphics[width=0.45\textwidth]{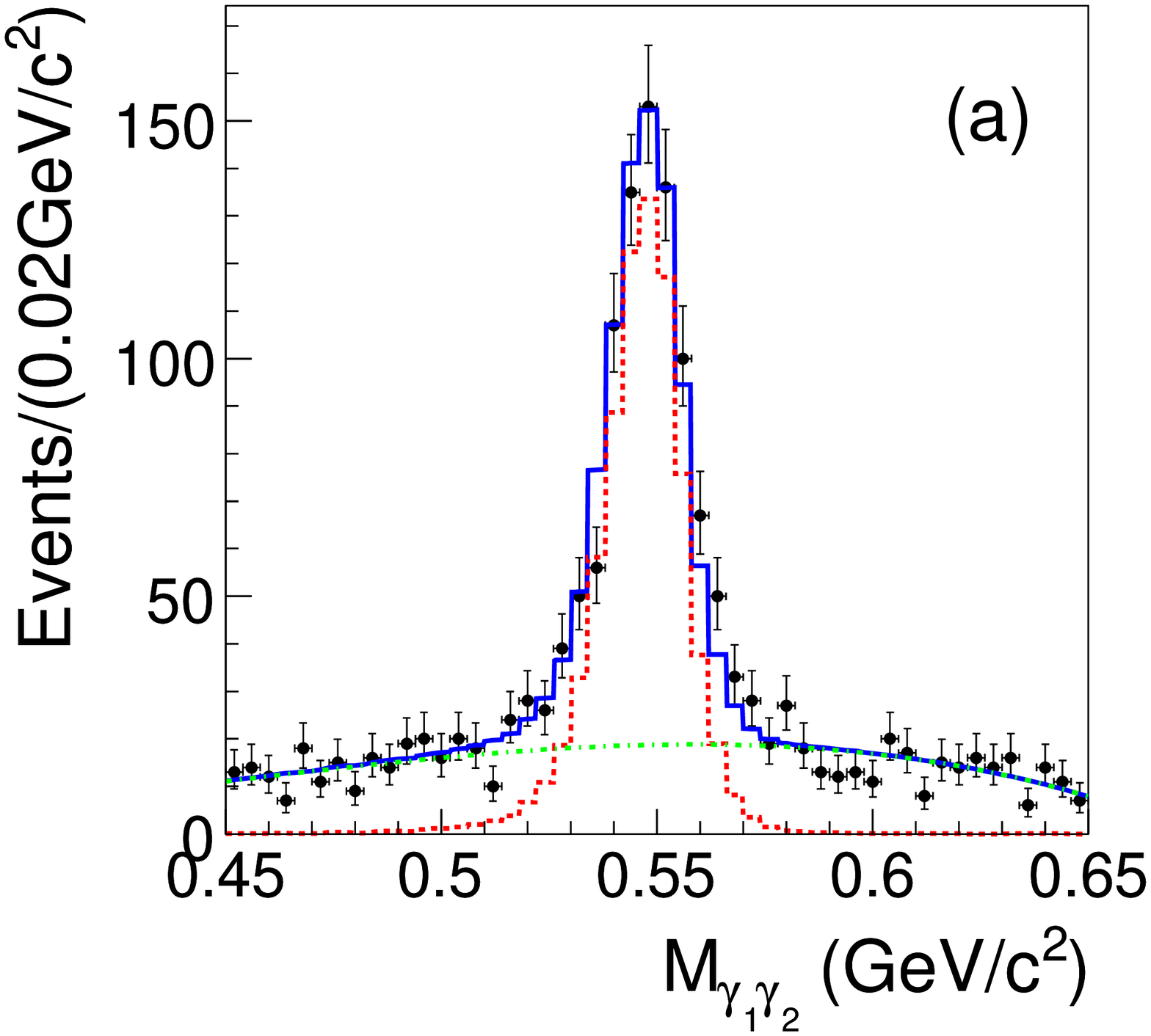}
\includegraphics[width=0.45\textwidth]{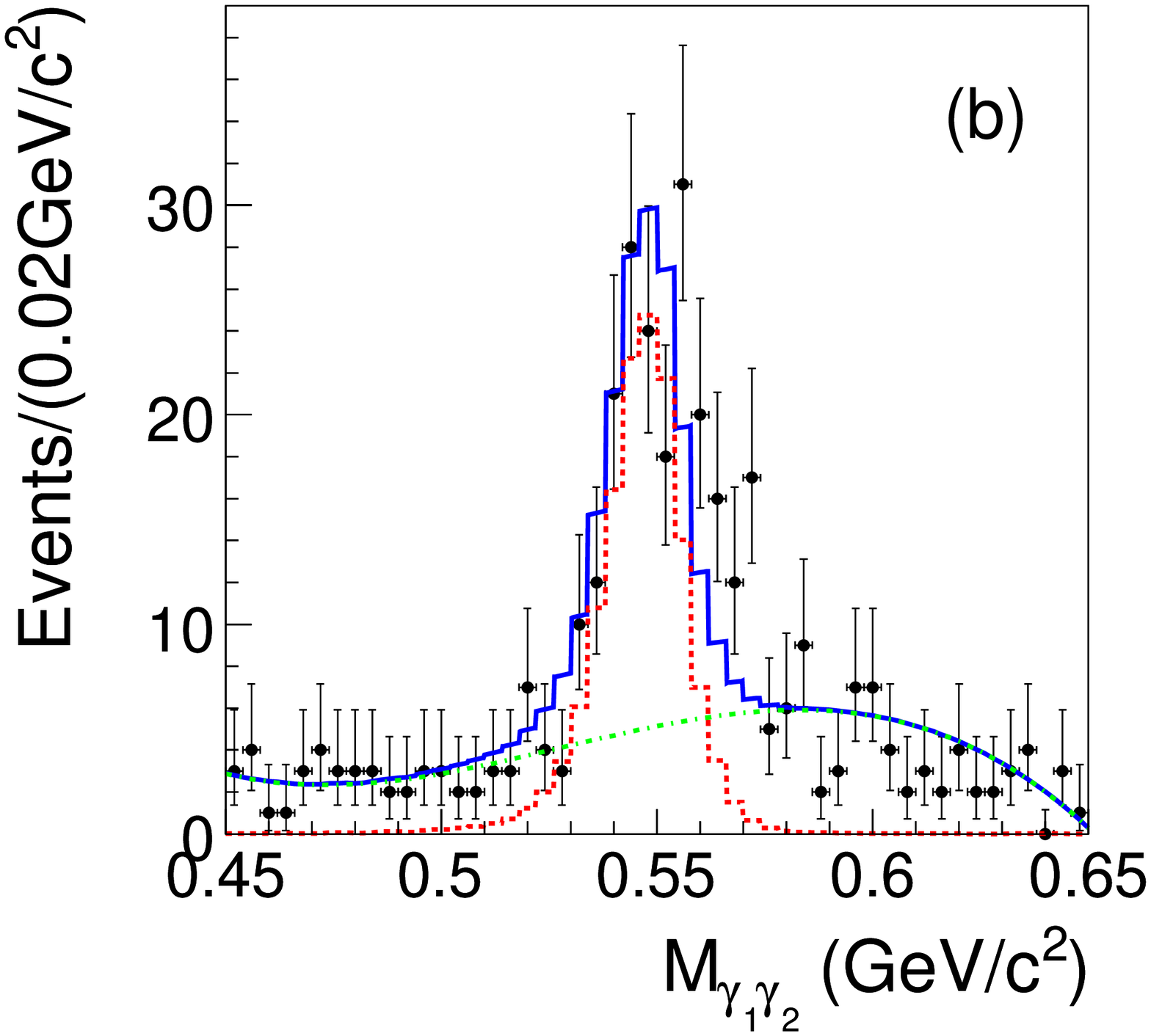}
\caption{(color online) Result of the fit to the $\eta$ mass distributions in the $\pi^{0}$ signal (a) and sideband (b) regions.
The circular dots with error bars show the distribution. The solid curve represents the fit result, while the short-dashed and dot-dashed curves represent
the $\eta$ signals and backgrounds, respectively.
\label{fig:data_fit_eta}}
\end{figure}

To obtain the number of $\gamma\eta\pi^0$ events, an unbinned maximum likelihood fit is performed to the mass spectrum of
the $\eta$ candidates, in the $\pi^{0}$ signal and sideband regions separately. The $\eta$ signals are parametrized by the
shape obtained from the signal MC. The background shape is described by a third order Chebychev polynomial.
The fit is shown in Fig.~\ref{fig:data_fit_eta}. The number of $\eta$ candidates obtained from the fit in
the $\pi^{0}$ signal region is $N = 746\pm34$, while in the $\pi^{0}$ sideband regions the corresponding number is
$N_\text{sideband} = 138\pm16$. The number of signal events is estimated to be
$N_\text{sig} = N - f\cdot N_\text{sideband} = 596 \pm 38$.

The number of peaking background events from
$J/\psi\rightarrow \omega\eta (\omega\rightarrow\gamma\pi^{0})$ and
$J/\psi\rightarrow \phi\eta (\phi\rightarrow\gamma\pi^{0})$
is obtained from exclusive MC samples, and the corresponding background yields
are given as $N_{J/\psi\rightarrow \omega\eta} = 122\pm4$ and $N_{J/\psi\rightarrow \phi\eta} = 16.5\pm0.1$.
The errors given here are the statistic errors from MC samples.

The $J/\psi\rightarrow\gamma\eta\pi^{0}$ branching fraction is calculated using the following expression:
\begin{eqnarray}\label{eq:BF}
\mathcal{B}(J/\psi \rightarrow \gamma\eta\pi^{0}) =
 \frac{N_\text{sig} - N_{J/\psi\rightarrow \omega\eta} - N_{J/\psi\rightarrow \phi\eta}}{N_{J/\psi} \times \mathcal{B}_{\eta} \times \mathcal{B}_{\pi^{0}} \times \varepsilon_\text{rec} }   ,
\end{eqnarray}
where $N_{J/\psi}$ is the total number of $J/\psi$ events~\cite{ref:jpsitotnumber},
and $\mathcal{B}_{\eta}$ and $\mathcal{B}_{\pi^{0}}$ are the branching fractions of the $\eta$ and $\pi^0$
decays to two photons, respectively~\cite{ref:pdg}.
The detection efficiency, $\varepsilon_\text{rec} = (24.5 \pm 0.2)~\%$, is obtained from the simulated signal events.
The resulting branching fraction is calculated to be $\mathcal{B}(J/\psi \rightarrow \gamma\eta\pi^{0}) = (2.14\pm0.18)\times10^{-5}$.

\begin{figure}[htbp]
  \centering
    \includegraphics[width=0.45\textwidth]{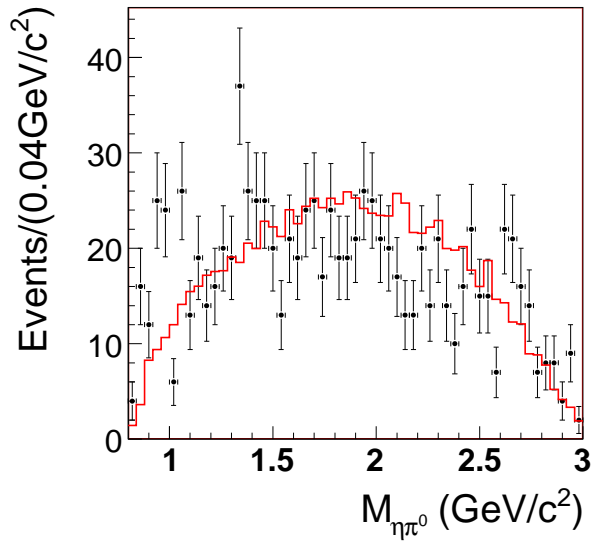}
\caption{Invariant $\eta\pi^{0}$ mass spectrum after final events
  selection and $\eta$ and $\pi^0$ mass cuts (points with error
  bars). The solid curve shows the phase space of $J/\psi \rightarrow
  \gamma\eta\pi^{0}$. }
  \label{fig:data_etapi0}
\end{figure}

We also investigate the intermediate resonant process $J/\psi \rightarrow \gamma X \rightarrow \gamma\eta\pi^{0}$, where $X$ stands for
$a_0(980)$ or $a_2(1320)$.
The $\eta\pi^{0}$ invariant mass spectrum in the $\eta$ and $\pi^{0}$ signal regions is shown in Fig.~\ref{fig:data_etapi0}. We perform an unbinned
maximum likelihood fit to determine the branching fractions of the radiative $J/\psi$ decays into these two mesons.
For the $a_0(980)$ signal shape, we use the $Flatt\acute{e}$ formula~\cite{ref:flatte} with the parameters from
the $K\bar{K}$ model~\cite{ref:WUjja0f0}, while the $a_2(1320)$ signal shape is described by a Breit-Wigner (BW) function
with the mass and width taken from PDG~\cite{ref:pdg}. The $a_0(980)$ and $a_2(1320)$ signal shapes are
convoluted with corresponding resolution functions, and multiplied by the efficiency distribution.
The resolution and efficiency as functions of the $\eta\pi^0$ invariant mass are obtained using the signal MC sample.
The resolution function is modeled by a sum of two Gaussians, with central values, widths and ratios fixed to the values obtained by analyzing
the mass resolutions of the $a_{0}(980)$ and $a_{2}(1320)$ resonances. The background shape consists of a third order Chebychev polynomial
and two functions obtained from MC study for the background channels $J/\psi\rightarrow\gamma\eta'$, $\eta'\rightarrow2\pi^{0}\eta$ and
$J/\psi\rightarrow\gamma\eta'$, $\eta'\rightarrow\gamma\omega$ .

\begin{figure}[htbp]
  \centering
    \includegraphics[width=0.45\textwidth]{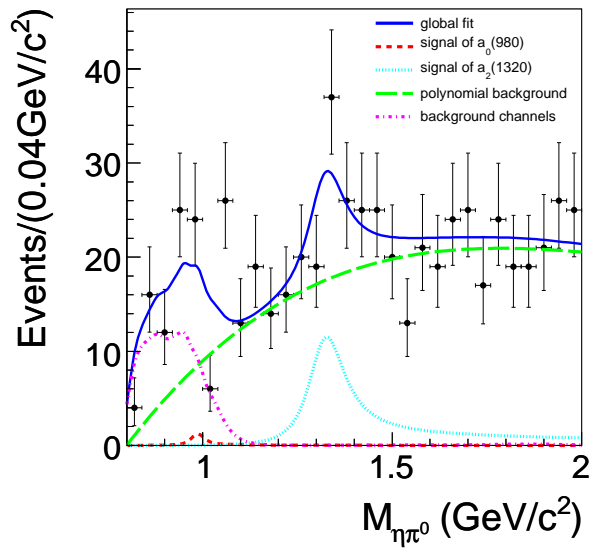}
\caption{(color online). Fit to the $\eta\pi^{0}$ mass spectrum in the [0.8, 2.0] GeV/$c^{2}$ region.
The points with error bars are data; the solid curve shows the overall fit projection; the short-dashed curve represents the $a_0(980)$ signal;
the dotted curve represent the $a_2(1320)$ signal; the dot-dashed curve corresponds to the two background channels $J/\psi\rightarrow\gamma\eta'$, $\eta'\rightarrow2\pi^{0}\eta$ and
$J/\psi\rightarrow\gamma\eta'$, $\eta'\rightarrow\gamma\omega$;  and
the long-dashed curve shows the remaining non-resonant $\eta\pi^{0}$ events.}
  \label{fig:upperlimit_fit_two}
\end{figure}

The spectrum in Fig.~\ref{fig:upperlimit_fit_two} is obtained from the fit to the first region,
[0.8, 2.0]~GeV/$c^{2}$. The event yields are $5$ for $a_0(980)$ and $57$ for $a_2(1320)$. The statistical significance is $0.5\sigma$ for $a_0(980)$ and $2.9\sigma$ for $a_2(1320)$.
Using a Bayesian method~\cite{ref:pdg}, we determine the upper limits for the $a_0(980)$ and $a_2(1320)$ production,
at the 95\% confidence level (C.L.), by finding the value $N_\text{sig}^\text{UL}$ such that
\begin{eqnarray}
\frac{\int^{N^\text{UL}_\text{sig}}_{0}\mathcal{L}dN_\text{sig}}{\int^{\infty}_{0}\mathcal{L}dN_\text{sig}}=0.95,\nonumber
\end{eqnarray}
where $N_\text{sig}$ is the number of signal events, and $\mathcal{L}$ is the value of the likelihood function of
$N_\text{sig}$ obtained in the fit. We find the upper limits at the 95\% C.L. on the number of
the $a_0(980)$ and $a_2(1320)$ to be $N^\text{UL}_{a_0(980)} = 26.0$ and $N^\text{UL}_{a_2(1320)} = 92.1$.

We study the upper limits under different assumptions for the shapes of the $a_0(980)$ and $a_2(1320)$  signal and non-resonant $\eta\pi^{0}$ processes.
For the non-resonant $\eta\pi^{0}$ process, we replace the third-order Chebychev polynomial with a fourth-order Chebychev polynomial
 or the $\eta\pi^0$ distribution from the signal MC. We also fit the signals of $a_{0}(980)$ and $a_{2}(1320)$
 together with background described above.
All these variations are applied in three different mass regions: [0.8, 2.0] GeV/$c^{2}$, [0.8, 1.92] GeV/$c^{2}$
and [0.8, 2.08] GeV/$c^{2}$. In addition, the fractions of the background channels are varied within one standard deviation due to
the MC statistics and the used branching fractions.
The signal shapes are varied by using different parameters of the $a_{0}(980)$ and $a_{2}(1320)$ functions.
In the $Flatt\acute{e}$ formula for the $a_0(980)$, the parameters from the $K\bar{K}$ model
are substituted by the $q\bar{q}$ model and $q\bar{q}g$ model parameters~\cite{ref:WUjja0f0, ref:prd90}.
In the case of the $a_{2}(1320)$, the mass and width of the BW function are varied within the uncertainties of the quoted values~\cite{ref:pdg}.
We take the largest upper-limit number of signal events among different models as a conservative estimate,
where we have the upper limits $N^\text{UL}_{a_0(980)} = 33.8$ corresponding to the $q\bar{q}g$ model,
while $N^\text{UL}_{a_2(1320)} = 107.9$ corresponding to a $1\sigma$ variation in the width for the $a_{2}(1320)$.

The upper limit on the product of branching fractions is determined by
\begin{eqnarray}\label{eqn2}
\mathcal{B}(J/\psi &\to& \gamma X, X\to \eta \pi^0) \nonumber\\
&<& \frac{N_{X}^\text{UL}}{N_{J/\psi} \times (1 - \sigma_\text{sys.}) \times \mathcal{B}_{\eta} \times \mathcal{B}_{\pi^{0}} \times \varepsilon }  ,
\end{eqnarray}
where $N_{X}^\text{UL}$ is corresponding number of signal events.
The efficiency is 16.7\% (20.1\%) for the $a_{0}(980)$ ($a_{2}(1320)$), obtained from the $J/\psi \rightarrow \gamma a_{0}(980)$ ($J/\psi \rightarrow \gamma a_{2}(1320)$) MC sample. $\sigma_\text{sys.}$ is the total systematic uncertainty of the quantities in the denominator in Eq.~(\ref{eqn2}).
The upper limits on the branching fractions are
$\mathcal{B}(J/\psi \rightarrow \gamma a_{0}(980) \rightarrow \gamma\eta\pi^{0}) < 2.5 \times 10^{-6}$ and
$\mathcal{B}(J/\psi \rightarrow \gamma a_{2}(1320) \rightarrow \gamma\eta\pi^{0}) < 6.6 \times 10^{-6}$ at the 95\% C.L.

\section{Systematic uncertainties}\label{sec:sys_error_br}

To estimate systematic uncertainties in our measurement of
the branching fractions,
we consider the following effects: photon detection efficiency, photon energy scale, photon energy resolution,
photon position reconstruction, the kinematic fit, and the fitting procedures.
Uncertainties associated with our fitting procedures stem from
the background shape, MC modeling of angular distributions, fitting region, background subtraction.
External factors include the total number of $J/\psi$ events, branching fractions of the intermediate states
and uncertainties in the branching fractions of the two background channels $J/\psi \to \omega\eta$ and $J/\psi \to \phi\eta$.

The systematic uncertainty from the photon detection is studied by comparing the photon detection efficiency between MC simulation
and a control sample consisting of the $J/\psi \rightarrow \rho\pi$ decays. The relative efficiency difference is about 1\% for
each photon~\cite{ref:photon1}. In this paper, 5\% is taken as the systematic error for the efficiency of
detecting five photons in the final state.

The uncertainty in the photon energy scale is determined to be 0.4\%~\cite{ref:bianjm}.
After varying photon energy according to this factor, we obtain the difference in the branching fraction of 1.9\%.

To estimate the uncertainty associated with the photon energy resolution, the photon energy is
smeared by the Gaussian with energy dependent width,  $\sigma_\text{smear} = 0.0083 \times E_{\gamma}$.
This factor is determined from the difference in relative energy resolution between data and MC of 4\%~\cite{ref:bianjm}.
With this smearing applied to the exclusive signal MC, we determine the corresponding efficiency and
find that the systematic error associated with the photon energy resolution is 0.9\%.

The difference in energy resolution between data and MC also affects the kinematic fit.
When we adjust the energy error in the reconstructed photon error matrix by 4\%~\cite{ref:bianjm},
we obtain a 1.1\% difference in the branching fraction measurement.

The uncertainty in photon position reconstruction is studied by changing the position parameter
of each photon in the signal MC and the difference is found to be negligible (less than 0.1\%).

When fitting two photons invariant mass distributions of the $\eta$ and $\pi^0$ candidate,
we vary the background shape by replacing a third order Chebychev polynomial with a second or fourth order polynomial.
The difference of 2.4\% with respect to our nominal result is associated with these effects.

The angular distributions of the $\eta$ and $\pi^0$ in the signal MC are based on the phase space model.
To obtain the uncertainty associated with this assumption, we change the angular distributions for
the $\eta$ and $\pi^0$ by assuming a form: $dN/d\cos\theta_{\eta/\pi^0}\sim (1+\cos\theta^2_{\eta/\pi^0})$.
We find the difference in the branching fraction of 9.2\% from this effect.

In the nominal fit, the mass spectrum of the $\eta$ is fitted in the range from
0.45 GeV/$c^{2}$ to 0.65 GeV/$c^{2}$. Alternative fits within ranges from 0.43 GeV/$c^{2}$  to 0.67 GeV/$c^{2}$ and
from 0.47 GeV/$c^{2}$ to 0.63 GeV/$c^{2}$ are performed, and the difference in the branching fraction of 1.6\% is taken as
the systematic uncertainty.

The uncertainty due to background subtraction is obtained by changing
the $\pi^{0}$ sidebands from $0.03$ GeV/$c^{2} <|M_{\gamma\gamma}- m_{\pi^{0}}| < 0.045$ GeV/$c^{2}$
to $0.035$ GeV/$c^{2} <|M_{\gamma\gamma}- m_{\pi^{0}}| < 0.05$ GeV/$c^{2}$, which corresponds to
a 1$\sigma$ change in sideband separation from the mass peak. The difference is found to be 2.0\%,
which is taken as the uncertainty from the background subtraction.

The number of $J/\psi$ events is determined from an inclusive analysis of the $J/\psi$ hadronic decays, and has an uncertainty of
0.6\%~\cite{ref:jpsitotnumber}. The uncertainties due to the branching fractions
of $\eta \rightarrow \gamma\gamma$ and $\pi^{0} \rightarrow \gamma\gamma$ are taken from PDG~\cite{ref:pdg}.
The uncertainties due to the branching fractions of the background channels
$J/\psi \rightarrow \omega\eta$ and $J/\psi \rightarrow \phi\eta$ are obtained by varying the respective values
within $1\sigma$~\cite{ref:pdg}.
The uncertainty associated with the branching fractions of background channels is determined to be 3.2\%.

All the contributions are summarized in Table~\ref{tbl:summary_sys_err}. The total systematic uncertainty is given
by the quadratic sum of the individual errors, assuming all sources to be independent.

\begin{table}
  \caption{Summary of systematic uncertainties(\%) in the measurement of the branching fractions. $\mathcal{B}_{a_{0}(980)}$ is the branching fraction of $J/\psi \rightarrow \gamma a_{0}(980) \rightarrow \gamma\eta\pi^{0}$ and
  $\mathcal{B}_{ a_{2}(1320)}$ is the branching fraction of $J/\psi \rightarrow \gamma a_{2}(1320) \rightarrow \gamma\eta\pi^{0}$. }
 \begin{tabular}{lccc}\hline\hline
 Sources & $\mathcal{B}(J/\psi\rightarrow\gamma\eta\pi^{0})$ & $\mathcal{B}_{a_{0}(980)}$ & $\mathcal{B}_{ a_{2}(1320)}$\\
 \hline
 Photon efficiency & 5.0 & 5.0 & 5.0 \\
 Photon energy scale &1.9 & 3.6& 3.8 \\
 Photon energy resolution &0.9 &0.6 & 0.5\\
 Kinematic fit & 1.1& 2.4 & 2.6\\
 Background shape & 2.4& - & - \\
 MC model & 9.2& - & -\\
 Fitting region &1.6 & - & - \\
 Background subtraction & 2.0 & - & - \\
 \hline
 Number of $J/\psi$ events & 0.6 & 0.6 &0.6\\
 Intermediate decays & 0.6 & 0.6& 0.6\\
$\mathcal{B}_{bg}$ & 3.2 & - & -\\
 \hline
 Total  & 11.8 & 6.7 & 6.9\\
\hline\hline
\end{tabular}
  \label{tbl:summary_sys_err}
\end{table}

\section{Summary}
Based on 223.7 million $J/\psi$ events collected with the BESIII detector,
the $J/\psi\rightarrow\gamma \eta \pi^0$ decay has been firstly observed.
The branching fraction of the $J/\psi\rightarrow\gamma \eta \pi^0$ process is measured to be
$(2.14\pm0.18\text{(stat)}\pm0.25\text{(syst)})\times10^{-5}$. With the Bayesian approach, upper limits for the
intermediate production of $a_0(980)$ and $a_2(1320)$ have been obtained at the 95\% C.L.
The upper limits are
$\mathcal{B}(J/\psi \rightarrow \gamma a_{0}(980) \rightarrow \gamma\eta\pi^{0}) < 2.5 \times 10^{-6} $ and
$\mathcal{B}(J/\psi \rightarrow \gamma a_{2}(1320) \rightarrow \gamma\eta\pi^{0}) < 6.6 \times 10^{-6} $, including systematic uncertainties.

For comparison, the branching fraction for the process
$J/\psi\rightarrow\gamma f_2(1270)\rightarrow\gamma\pi^{0}\pi^{0}$ is $(4.0\pm0.09\pm0.58)\times10^{-4}$~\cite{ref:zhuyc}, while for $J/\psi\rightarrow\gamma f_0(1500)\rightarrow\gamma\pi^{0}\pi^{0}$ it is $(0.34\pm0.03\pm0.15)\times10^{-4}$~\cite{ref:zhuyc}.
This study shows that the suppression rates for isospin-one processes in $J/\psi$ radiative decays,
compared to isospin-zero decays, are consistent with naive theoretical expectations~\cite{jpsidecay:1989}, \emph{i.e.}, at least one order of magnitude. It is noticed that the upper limit on
$\mathcal{B}(J/\psi \rightarrow \gamma a_{0}) \times  \mathcal{B}(a_{0} \rightarrow \eta\pi^{0})$ is much lower than the theoretical calculation in Ref.~\cite{ref:theory}.
The result in this paper indicates that the decay mechanism of $J/\psi \rightarrow \gamma a_0(980)$ may be totally different from
$\phi \rightarrow \gamma a_0(980)$, so the factorization method may not work for the $J/\psi \rightarrow \gamma a_0(980)$ decay~\cite{ref:theory}.
Our measurement provides important constraints on theoretical calculations.

\section{\bf ACKNOWLEDGMENTS}
The BESIII collaboration thanks the staff of BEPCII and the IHEP computing center for their strong support. This work is supported in part by National Key Basic Research Program of China under Contract No. 2015CB856700; National Natural Science Foundation of China (NSFC) under Contracts Nos. 11125525, 11235011, 11322544, 11335008, 11425524; the Chinese Academy of Sciences (CAS) Large-Scale Scientific Facility Program; the CAS Center for Excellence in Particle Physics (CCEPP); the Collaborative Innovation Center for Particles and Interactions (CICPI); Joint Large-Scale Scientific Facility Funds of the NSFC and CAS under Contracts Nos. 11179007, U1232201, U1332201; CAS under Contracts Nos. KJCX2-YW-N29, KJCX2-YW-N45; 100 Talents Program of CAS; National 1000 Talents Program of China; INPAC and Shanghai Key Laboratory for Particle Physics and Cosmology; German Research Foundation DFG under Contract No. Collaborative Research Center CRC-1044; Istituto Nazionale di Fisica Nucleare, Italy; Koninklijke Nederlandse Akademie van Wetenschappen (KNAW) under Contract No. 530-4CDP03; Ministry of Development of Turkey under Contract No. DPT2006K-120470; National Natural Science Foundation of China (NSFC) under Contracts Nos. 11405046, U1332103; Russian Foundation for Basic Research under Contract No. 14-07-91152; The Swedish Resarch Council; U. S. Department of Energy under Contracts Nos. DE-FG02-04ER41291, DE-FG02-05ER41374, DE-SC0012069, DESC0010118; U.S. National Science Foundation; University of Groningen (RuG) and the Helmholtzzentrum fuer Schwerionenforschung GmbH (GSI), Darmstadt; WCU Program of National Research Foundation of Korea under Contract No. R32-2008-000-10155-0.

\end{document}